\documentclass[twocolumn,letterpaper,aps,prb,floats,floatfix]{revtex4}
\usepackage{amsmath}
\usepackage{psfig}
\usepackage{longtable}
\begin{document}
\bibliographystyle{prsty}
\title{Chemical Reaction Dynamics within Anisotropic Solvents in
         Time-Dependent Fields}
\author{Eli Hershkovits}
\thanks{Present address: Department of Electrical and
 Computer Engineering, Georgia Institute of Technology, Atlanta GA} 
\email{eli@bme.gatech.edu}
\author{Rigoberto Hernandez}
\affiliation{%
Center for Computational Molecular Science and Technology\\
School of Chemistry and Biochemistry\\
Georgia Institute of Technology\\
Atlanta, GA 30332-0400}
\email{heernandez@chemistry.gatech.edu}
\date{\today}
\begin{abstract}
The dynamics of low-dimensional Brownian particles
coupled to time-dependent driven anisotropic heavy particles
(mesogens) in a uniform bath (solvent) have been described
through the use of a variant of the stochastic Langevin equation.
The rotational motion of the mesogens is assumed to follow the
motion of an external driving field in the linear response limit.
Reaction dynamics have also been probed using a two-state
model for the Brownian particles.
Analytical expressions for diffusion and reaction rates have been
developed and are found to be in good agreement with numerical calculations.
When the external field driving the mesogens is held at 
constant rotational frequency,
the model for reaction dynamics predicts that the applied field frequency
can be used to control the product composition.

\end{abstract}
\maketitle 

\section{Introduction}

The stochastic or Brownian motion of a particle in a uniform solvent
is generally well-understood.\cite{risken89,rmp90}
The dynamics is less clear
when the solvents respond in a non-uniform or time-dependent manner,
although such problems are not uncommon.  
For example, the dynamical properties of 
 a suspension in a liquid crystal can be projected onto 
 an anisotropic stochastic equation of motion.\cite{yang,widata}
Other examples may
 include diffusion and reaction in supercritical liquids,\cite{tucker01}
 liquids next to the liquid vapor critical 
 point\cite{tucker01b,harris02,tucker02R} and growth in living
 polymerization.\cite{hern99a}

The flow properties of liquid crystals have generally been
  analyzed from the
  perspective of macroscopic nematohydrodynamics.\cite{deggen93} 
Therein, liquid crystals have been classified according
 to the presence or absence of solvent.
 Pure liquid crystals containing no solvent are called thermotropic
 in part because they have exhibited strong temperature-dependent
 behavior.
 A suspension of nematogens (anisotropic molecules)
  within a simple solvent is known as a lyotropic liquid.
The presence of nematogens leads to different transport properties within
the solvent than would be seen in a pure simple liquid alone.
  The additional complexity is a result of the coupling between
the velocity field and the average direction of the nematogens.
  As a result, the dynamics of a particle in the liquid crystal
is dissipated by a friction whose form is that of
a tensor and not a scalar.\cite{diogo}
  The actual drag can be further complicated
  by the presence of topological discontinuities in the liquid.\cite{ruhter}
To our knowledge, analytic solutions for the diffusion of Brownian particles 
in these general environments are not known.  
The situation for a reactive solute is even less
clear as no analytic formalism has been constructed.
In the present work, we construct a formalism---that in some 
limits---fills in these gaps.

%

One step toward understanding the dynamics in anisotropic
liquids would thus be the development of a lyotropic model 
consisting of a Brownian particle in the presence of a 
time-dependent driven mesogen.\cite{herher01}
Another step toward this goal is the analytic and/or numerical
solution of such.
In the present work, the rigorous construction necessary
for the first of these steps is not attempted.
Instead, a naive phenomenological model 
describing the dynamics in lyotropic liquids has been 
constructed.
It serves as a benchmark 
for the development of techniques useful in analyzing
the dynamics of Brownian particles dissipated by 
an anisotropic solvent through a time-dependent friction.
In particular, the
  lyotropic liquid is assumed to be nematic,
{\it i.e.}, the (calamitic) mesogens are assumed to be rod-like
as is the case with mineral moieties in water\cite{chrpat}.
The mesogens are further assumed 
  to be one-dimensional and rigid, and a series of additional 
  simplifying assumptions have been invoked. 
 A physical system rigorously
 satisfying all these assumptions may not exist,
but the benchmark
 may still exhibit some of the important dynamics that has been seen
 in real liquid crystals in the presence of magnetic fields
 with time and space instabilities.\cite{sramey89}
Another step toward understanding the dynamics in anisotropic
liquids is the rigorous solution of a thermotropic (nematic) model 
in which the
dilute Brownian particle diffuses or isomerizes in a solvent
that consists exclusively of mesogens.
  It is based on the possible connection to a rotating nematic liquid system 
  previously observed,\cite{brochard74,meyer93} 
and on the analytic understanding of the dynamics in nematic liquids 
in a few special cases.\cite{diogo,roman89}
  For this thermotropic case, 
  we don't attempt to develop a microscopic model of the
  friction and instead
  make assumptions based on the known properties of isotropic liquids.

  In general, the complicated microscopic dynamics of a subsystem coupled
  to a many-dimensional isotropic heat bath can be 
  projected onto a simple reduced-dimensional stochastic equation of motion
  in terms of the variables of the subsystem alone.
In the limit when
  the fluctuations in the isotropic bath are uncorrelated,
  the equation of motion is the Langevin Equation (LE),\cite{risken89}
\begin{subequations}\label{eq:le}\begin{eqnarray}
   \dot q &=& p \\
   \dot p &=& -V^{'}(q)-\gamma p +\xi(t) \;.
\end{eqnarray}\end{subequations}%
where $(q,p)$ are the position and momenta vectors
  in mass-weighted coordinates ({\it i.e.} mass equals one), $V(q)$ 
  is the system
  potential, $\gamma$ is the friction and $\xi$ is a 
  Gaussian random force due to the thermal bath fluctuations. 
  The friction and the random force
  are connected via the fluctuation dissipation theorem,
  \begin{equation}
  \label{eq:fdr}
   \langle\xi(t_1)\xi(t_2)\rangle
        = {2\gamma\over\beta}\delta(t_1-t_2)\;,
  \end{equation}
where the average is taken over all realizations
of the forces at the inverse 
  temperature $\beta\,[\equiv (k_{\rm B}T)^{-1}]$.
  The LE can represent the generic problem of the
  escape rates of a thermally activated particle from a 
  metastable well when the thermal energy is much lower
  then the barrier height.\cite{rmp90}
  The one-dimensional LE has been solved in the asymptotic limits of 
  weak and strong friction by Kramers.\cite{rmp90}
  A general solution for weak to intermediate friction
  was found by Melnikov and Meshkov.\cite{mm86}  This result was 
subsequently
  extended to the entire friction range in the turnover theory of
Pollak, Grabert and H{\"a}nngi.\cite{pgh89}
  The reactive rates for a multidimensional LE
have been obtained exactly in the
  strong\cite{lang69,schuss82,schuss83,berne84,bork85,nitzan88} 
  and weak friction\cite{berne84} limits and 
approximately in between these limits through
a multidimensional turnover theory.\cite{herpol97} 
  The LE can also describe the dynamics of a subsystem under 
  an applied external force,
and has led to the observation of such 
  interesting phenomena as 
  stochastic resonance,\cite{bsv81,march98}
  resonant activation,\cite{DG92,RH97} and rectified 
  Brownian motion.\cite{manman00,mag93,Reim02,Reim03}
%

  When the fluctuations in the 
isotropic 
  bath do not decay quickly in space or in time,
the dynamics are known to be described by the
  the Generalized Langevin Equation (GLE).\cite{grot80}
The activated rate expression
  for a particle described by a GLE is also well-known.\cite{hang82,carm82}
Less understood are the exact rates when 
the friction 
 dissipates the subsystem differently at different times
 in a nonstationary GLE-like 
 equation.\cite{tucker01,hern99a,hern99b,herher01,hern99d,haynes94}
Nonetheless, the models developed in this work contain
the flavor of this nonstationarity in that 
the LE is driven by an
  external periodic field through the friction rather 
than through a direct force on the system.
  Consequently the result of this study also provide new insight
  into the dynamics of systems driven out of equilibrium.

The primary aim of the paper is the
development of analytical and numerical techniques to 
obtain the diffusion and reaction rates of
a subsystem dissipated by a time-dependent driven anisotropic
solvent in various limits. 
A naive model for a nematic lyotropic liquid and its various underlying
assumptions is presented in Sec.~\ref{sec:model}
as one paradigmatic example for the accuracy
of the methods described in this work. 
Another model based on an experimental
system of the rotating nematic liquid is 
described briefly in Sec.~\ref{sec:nematic}.
The anisotropic solvent is manifested in these models by way of
a time-dependent friction that is externally driven.
The diffusion of
 free Brownian particles
dissipated by a time-dependent environment is described in 
Sec.~\ref{sec:free}. 
 The numerical methods for calculating reactions rates
needed to extend the solutions of these models to 
include nontrivial potentials of mean force are presented
in Sec.~\ref{ssec:numerical}.
 Analytical approximations for 
otherwise-rigorous rate formulas are derived and compared to the 
the numerical results in Sec.~\ref{ssec:analytic}. 
 A discussion of the validity of all of these approaches and
possible applications concludes the paper in Sec.~\ref{sec:conc}.

\section{A naive lyotropic model with rotating external fields}
\label{sec:model}

A naive model describing a particle propagated in an
anisotropic solvent is motivated in this section in the 
context of diffusive or reactive dynamics within 
a lyotropic solvent.
The connection between the model and realizable lyotropic solvents
is only a loose one. No attempt is made here to do
a rigorous projection of the detailed complex modes 
of the lyotropic solvent onto the subsystem dynamics.
The lyotropic liquid is assumed to consist of
rod-like mesogens and a uniform isotropic liquid solvent.
It is further assumed that there exists a single tagged motion
characterized by an effective coordinate $q$
that describes the subsystem 
---{\it e.g.}, a probe particle or reacting pair of particles---
whose dynamics is of interest.
This tagged motion is taken to be one-dimensional for simplicity.
The effective mass $m_q$ associated with the tagged subsystem is 
also assumed to be well separated from the smaller mass of the
isotropic liquid, and the larger mass of the (anisotropic)
rod-like mesogens.
Consequently the tagged motion can be described as that of a
Brownian particle at position $q$ experiencing a dissipative environment
due to the interactions with the isotropic liquid and the mesogens.



The model is further simplified by assuming that the mesogens
of given concentration, $c$,
do not interact with each other.
This ideal-solute assumption is certainly realized at low
enough concentrations that the mean spacing between mesogens is long compared
to their effective interaction distance.
(It would be easy to achieve such concentrations
even at relatively high concentrations if the interaction
potentials are hard-core.)
  The ideal-solute mesogens will exhibit no orientational
order in the absence of external fields. 


In real nematic liquids there are 
  interactions between the mesogens
  that result from cooperative forces. 
They, as well as boundary effects on the rods,
are excluded within the model of this work.
  The orientation of all the rods is firmly fixed by a magnetic field 
(homogeneous director field) with inclination
  $\theta$ relative to the $y$ axis:
 \begin{subequations}\label{eq:Hfield}\begin{eqnarray}
   H_x&=&H_0\sin{\theta} \\ 
   H_y&=&H_0\cos{\theta}\\
   H_z&=&0 \;,
 \end{eqnarray}\end{subequations}%
  This strong field assumption ---all the mesogens will orient 
uniformly in the direction of $\vec H$---
also ensures that there is no angular momentum
  transfer in collisions 
between the mesogens and diffusing Brownian solutes.
The environment is clearly anisotropic, and a Brownian particle
diffusing through it would experience different dissipative
forces depending on the direction of its motion.
\begin{figure}
\centerline{\psfig{figure=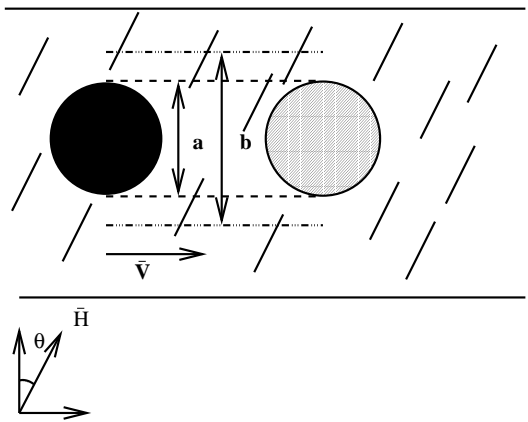,width=2.75in}}
\caption{A Brownian particle with a diameter $2a$ moves with
 velocity $\vec v$ inside a mixture of an isotropic liquid 
 and calamitic mesogens.
  The mesogens have a length $l$ of the same order of $a$,
  a negligible width and concentration, $c$. 
  The mesogens are oriented by an external magnetic field $\vec H$. 
  The magnetic field is characterized by the angle $\theta$ 
relative to $\vec v$.
  Under these conditions, the Brownian particle collides with 
  $(\pi R^2+2Rl|\cos\theta|)$ mesogens per unit time.}
\label{fig:cartoon}
\end{figure}
The suspended particle is assumed to have a spherical shape with a radius $R$.
The particle velocity $v(t)$ is restricted to the $x$ direction.
The number of collisions per unit time between the Brownian particle
and the mesogens is simply 
$(\pi R^2+2Rl|\cos\theta|)vc$.
This result is illustrated in Fig.~\ref{fig:cartoon}.
Further assuming that each of the mesogens has 
a thermal distribution of velocities and 
noting the mass difference between the mesogens and the environment,
 the friction force on the Brownian particle 
 is proportional to the number of collisions. 
 This gives a friction coefficient:
  \begin{equation}
   \gamma=\gamma_0(\pi R^2+2Rl|\cos\theta|)\;, \label{eq:gamma}
  \end{equation}
  where $\gamma_0$ is a proportionality constant characteristic of 
the system.
The viscosity of the isotropic liquid leads to
  additional dissipation that is manifested as an additional
  constant term to the overall friction.
However, in those cases when this isotropic friction is 
dominated by the friction of Eq.~\ref{eq:gamma}, its effect is
small and insufficient to blur the anisotropy of the system.
For further simplicity, therefore, in what follows, the isotropic friction
due to the homogeneous solvent will be assumed to be zero 
without loss of generality as long as the actual isotropic friction is 
weak in this sense.

  The instantaneous inclination $\theta(t)$ has a large influence on
  the short-time dynamics of a particle whose motion is 
measured only along an
  initially chosen $x$ axis.
Without this restriction, a fixed $\theta$ 
  will not influence the dynamics 
because the particle motion will necessarily average
  over all directions.  As a result, the inclination can be used as a control
  parameter when one
measures only the dynamics along
  a specific direction
  but not when one is interested in the average diffusion or
  reaction of the chosen subsystem with respect to all directions.

In cases when the magnetic field of Eq.~\ref{eq:Hfield}
rotates with frequency $\omega$, then 
  the Brownian particle experiences a friction,
  \begin{equation}
   \gamma(t)=\gamma_0(\pi R^2+2Rl|\cos\omega t|)\;,
\label{eq:gammaoft}
  \end{equation}
 that is periodic in time.
  Including the dissipation of the rotating mesogens will not change this 
  friction, but
  will add a finite temperature to the bath due to rotational dissipation.
As long as 
  this amount of heat is much smaller then the bath temperature,
  the friction in Eq.~\ref{eq:gammaoft}
  is well-defined and can be used as the friction entirely
  dissipating the Brownian particle.

\section{A nematic model with external rotating fields} \label{sec:nematic}

While the naive model described above does capture some of the
features of liquid crystal diffusion, it is nonetheless too
simplistic.
%
Experiments of pure nematic liquids 
  under a rotating magnetic field\cite{brochard74,meyer93}
can serve to illustrate the possibility of solvent responses characterized
by
  time dependent viscosity.
  In these experiments,
the homogeneous director field
  of a nematic liquid confined between two parallel glass plates
was aligned in the plane of the plates by strong magnetic field.
The magnetic field was also rotated at constant velocity within this plane.
For many of the experimental conditions, the
nematic liquid retained uniform alignment but its homogeneous director field
followed the magnetic field with a constant phase lag.
  Finding an expression for the viscosity in a nematic liquid is far 
  more complicated then for an isotropic liquid. It has five coefficients
\cite{deggen93} and depends on the orientation of the director, the velocity
  and the velocity gradient. This problem was only partially solved for some 
special cases.  One case obtains the effective viscosity in a suspension
  of small particles in a nematic liquid.\cite{roman89}
The key simplifications are that the small particles are assumed to 
  be not much larger than the nematogens and with spherical shape.
The friction coefficient has the simple form,
  \begin{equation}
   f_i=a(A \delta_{ik}+B\cos^2\theta)\;,\label{eq:caseone}
  \end{equation}
  where the expression for constant coefficients $A$ and $B$ may be found in 
Ref.~\onlinecite{roman89}.
A second case treats the limit in which the chosen particle in a
  nematic liquid has a large spherical shape.\cite{diogo}
The resulting effective friction is composed of an isotropic term
  and an anisotropic term that depends on the angle between the director
  and the particle velocity.
  The anisotropic expression is a little more 
  complicated than Eq.~\ref{eq:caseone},
  but its leading order terms also involve $\sin\theta$ and $\cos\theta$.
  In both of these cases, the nematic liquid is assumed
  to be firmly oriented by a strong external field and the friction
  force is taken to be much larger than the elastic forces in the nematic 
  liquid. 
Thus the naive model described in the previous section does exhibit 
both a uniform constant term and an anisotropic oscillatory term that 
are in qualitative---though not quantitative---agreement with more
detailed models.

\section{Free Brownian Diffusion in an Anisotropic Solvent}\label{sec:free}

The
 motion of a free Brownian particle in the time-dependent friction 
 field of Eq.~\ref{eq:gammaoft} can
 be described by the Langevin equation,
 \begin{equation}
  \dot{p}=-\gamma_0\phi(t)p+\psi(t)\xi(t) \;,
\label{eq:tlan}
 \end{equation}
  where the time-dependent coefficients,
\begin{subequations}\label{eq:tfric}\begin{eqnarray}
    \phi(t)&=&\psi(t)^2\\ 
    \psi(t)&=&a+\cos(\omega t) \;,
\end{eqnarray}\end{subequations}%
have been chosen to describe 
  the periodic behavior of the naive lyotropic model 
  and the hydrodynamical friction terms in pure nematics
as simply as possible.
  The noise is related to the friction by the fluctuation dissipation
  relation,
  \begin{equation}
   \langle\xi(t)\xi(t')\rangle={2\gamma_0\over{\beta}}\delta(t-t')\;.
  \end{equation}
The strength $a$ of the isotropic term has been chosen to be
$1.05$ throughout the illustrations in this work 
to emphasize the anisotropic effects, but different 
physically-realizable strengths 
do not lead to different conclusions.
%

  The solution to the equation of motion \ref{eq:tlan} is
  \begin{eqnarray}
   p(t)&=&p_0 \exp[-\gamma_0G(t)]
\nonumber\\
       &+&\int_{t_0}^t dt_1\psi(t_1)\xi(t_1)
      e^{ -\gamma_0\{G(t)-G(t_1)\}} \;,
  \end{eqnarray}
where $p_0$ and $t_0$ satisfy the initial condition, $p_0\equiv p(t_0)$, and
  the integrated friction $G(t)$ is defined as
  \begin{equation}
   G(t)=\int_{t_0}^t dt_1\psi(t)^2 \;.\label{eq:ifric}
  \end{equation}
  The velocity correlation function is readily calculated to be
  \begin{eqnarray}
   \langle p(t_1)p(t_2)\rangle
      &=&{1\over{\beta}}
    \exp{\left[ -\gamma_0\{ G(t_1)+G(t_2) \right.}
       \nonumber\\&&\left.
      -2G(\min(t_1,t_2))
                       \}\right] \;.
  \end{eqnarray}
  The square mean displacement of the free particle after time t is the double 
  integral,
\begin{subequations}\label{eq:q2}\label{eq:13}\begin{eqnarray}
   \langle (q(t)&-&q(t_0))^2\rangle
\nonumber\\
        &=&\int_{t_0}^t\int_{t_0}^t\!dt_1dt_2\, \langle p(t_1)p(t_2)\rangle \\
     &=& {1\over{\beta}}
         \left[\int_{t_0}^t dt_1\int_{t_0}^{t_1}dt_2
               e^{[-\gamma_0\{G(t_1)-G(t_2)\}]}
         \right.\nonumber\\ 
      && \left.
         +\int_{t_0}^t dt_1\int_{t_1}^tdt_2
               e^{ [\gamma_0 \{G(t_1)-G(t_2)\}]} \right]
   \;.
\end{eqnarray}\end{subequations}%
 A similar expression was developed by Drozdov and Tucker\cite{tucker01}
 for the case of fluctuations in the local density of 
 supercritical solvent.
 The result in 
Eq.~\ref{eq:q2} leads to
 the diffusivity of the particle.
As will be shown below,
 the diffusivity in the
 time-dependent environment deviates from the linear correlation 
known to result
 in the constant friction environment.
  
\begin{figure}
\centerline{\psfig{figure=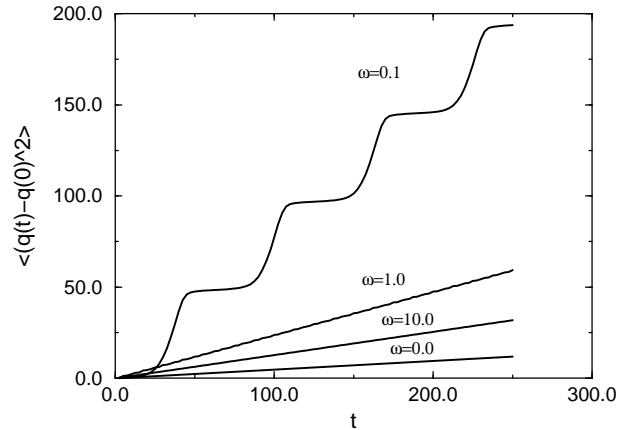,height=2.5in,angle=-90}}
\caption{The mean square displacement of a free Brownian particle in the 
naive lyotropic bath model has been obtained by direct integration
and through the use of the analytical expression 
in Eq.~\protect\ref{eq:13} at various frequencies $\omega$ of the 
driving rotating magnetic field.
In the former integration method, 
100,000 trajectories were sufficient to obtain convergence.
In the latter,
  the average is taken over an ensemble of 100,000 particles
  starting at time $t=0$ with inclination perpendicular to the velocity,
and overlays the results of the former within the resolution of the figure.
}
\label{fig:qtq0}
\end{figure}
  In Fig.~\ref{fig:qtq0},
  the mean square displacement 
of a Brownian particle whose motion is measured only along an
arbitrary one-dimensional axis
  is plotted as a function of time 
at various applied
  frequencies $\omega$.
  The 
{\it average behavior} 
  of the mean square displacement is linear with time,
as in the constant friction regime, but it also contains
  fluctuations (in time)
  around the average 
whose frequency depends on the external field.
It is important to note that the overall slope of the mean 
square displacement depends on the frequency $\omega$; that is,
  the diffusivity shows strong dependence
  on $\omega$. 
  Hence by changing the frequency of the external field, it becomes
  possible to control the diffusivity of the Brownian particles.

  The analytical result of Eq.~\ref{eq:q2} was used 
  to check the accuracy of the numerical 
  integrator 
employed in propagating
  particles in a time-periodic white noise bath. 
  A fourth-order
  integrator was developed based on the Taylor method in 
  Refs.~\onlinecite{herher01} and \onlinecite{art2}
  and is outlined in Appendix \ref{sec:integrator}.
  Such an algorithm is extremely useful 
as a check for
  nonstationary problems in which the
  integration time can be very long. 
  The new algorithm agrees with the analytical
  result up to time steps of size, $\delta t =.5$, 
in the dimensionless units of time defined in Eq.~\ref{eq:tlan}.
In general, the time step required to achieve a given accuracy
decreases as either the frequency or friction increases.

  These results are limited to diffusion in one dimension. When studying the 
  motion in the plane defined by the rotating magnetic field an average has
  to be taken over all the directions.
  The integrated friction function, Eq.~\ref{eq:ifric}, for a particle with the initial velocity
  inclined with angle $\omega\tau$ 
relative to the 
  magnetic field at the time, $t=0$, is
   \begin{eqnarray}
   G(t+\tau)&=&\gamma_0\left[a^2t+{2a\over{\omega}}\sin(\omega(t+\tau))
\right.\nonumber\\
     &&\left. +{t\over 2}+
             {\sin[2\omega(t+\tau)\over{4\omega}]}\right]\;.
\label{eq:Gconst}
  \end{eqnarray}
  After some elementary algebra, the integration in \ref{eq:q2} 
with $G$ as in Eq.~\ref{eq:Gconst}
  for the case of a constant magnetic field leads to the
  average diffusion coefficient,
  \begin{equation}
    \langle D\rangle_0={1\over{\gamma_0\beta}}{a\over{(a^2-1)^{3/2}}}\;,
\label{eq:15}
  \end{equation}
of a Brownian particle in a plane.
  The diffusivity 
of the Brownian particle 
  in a rotating field at various frequencies
\begin{figure}
\centerline{\psfig{figure=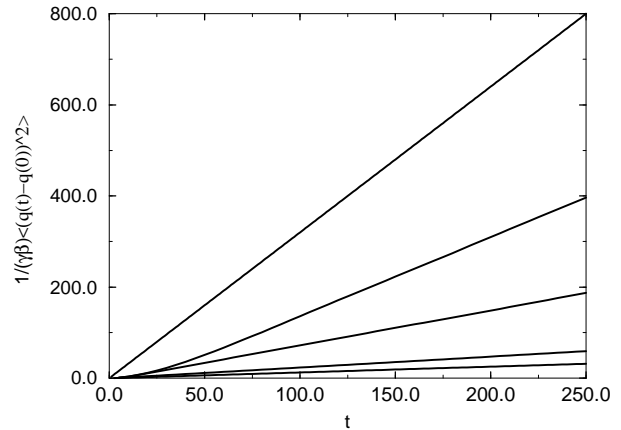,height=2.5in,angle=-90}}
\caption{The normalized average displacement of an ensemble of
free Brownian particles in the presence of a periodic friction
is displayed as a function of time.
The driving frequencies are $\omega=10, 1, 0.1, 0.2$ and $0$.
The result for the fixed case ($\omega=0$) has been calculated analytically
  using Eq.~\protect\ref{eq:15}.
The remaining results are obtained numerically by averaging
  over Brownian particles
  with velocities in random inclination relative to
  the magnetic field at the initial time, $t=0$. 
Note that  
the slopes ---{\it viz.}, the diffusion rate---
increase with decreasing $\omega$.}
\label{fig:diffusion}
\end{figure}
has been obtained numerically and is shown in Figure \ref{fig:diffusion}.
  As can be seen, the diffusivity is a monotonic decreasing function of the 
  frequency.
This result suggests the use of the applied field frequency to control
 the diffusive transport of the Brownian particle.

In the $a=1$ limit of this model, there is a divergence in 
   the averaged diffusion constant over all the directions 
   at constant magnetic field.
This limiting behavior is a consequence of 
   a transition from diffusive to ballistic motion at
   the inclination in which $\theta=\pi$. 
That is to say, that it is an artifact of the model in so far as the 
   physical system it represents would never take on the value of
   $a=1$, and hence would not exhibit an infinite diffusion!
Nonetheless, the model above serves to demonstrate the
   the accuracy of the numerical and analytical formalism
   when $a$ is far from 1.

\section{Reaction Rates in a Time-Periodic Friction}\label{sec:trates}

  The assumptions introduced in 
Sec.~\ref{sec:model} are also applicable to the description
of the reactive interactions between two
  Brownian particles. 
  Neglecting the hydrodynamic interaction 
as before,
  the dynamics can be described by the time-dependent Langevin
  equation,
  \begin{eqnarray}
   \ddot{q}=-\nabla V(q)-\gamma_0\phi(t)\dot{q}+\psi(t)\xi(t)\;,
\label{eq:dwpot}
  \end{eqnarray}
where
  $q$ is now a relative mass-weighted
coordinate between the interacting particles,
and
  $V$ is the potential of mean force between the particles.
  The remaining symbols are the same
  as in the previous section.
Phenomenological rate constants
---{\it e.g.}, transition from one metastable state of the potential to
  another or to infinity--- 
  cannot be calculated analytically when the potential
  is of a more complex form than that of the harmonic oscillator.
  Direct numerical 
evaluation of these rates
  is usually quite time consuming because the time
  scales involved in the problem are widely varying.
The reactive flux method reduces much of the computation
time by initiating the trajectories at the barrier.\cite{rmp90}
It has been used to obtain reactive exact thermal
escape rates in the stationary limit both numerically and exactly, and
to obtain approximate rates
under a variety of limiting approximations.
In the present case, the problem is nonstationary at short times but retains an
average stationarity at sufficiently long times. 
The strategy is consequently to generalize the 
rate formula for stationary systems.
It must now include processes in which stationarity is required only 
when the observables are integrated over a period 
equal to that of the external periodic perturbation.

In all of the calculations performed here to illustrate the approach, 
the potential has been chosen to have the 
the form of a symmetric quartic potential,
  \begin{equation}
  \label{eq:quar}
   V(q)=q^4-2q^2\;,
  \end{equation}
in which the two minima represent two distinct metastable states
separated by a dimensionless barrier of unit height.
(Note that for simplicity, all observables in this work are written
in dimensionless units relative to the choice of this effective
barrier and the particle mass.)
  The reactive rate has been calculated for particles with 
inverse temperatures, 
  $\beta= 10$, or $20$.
  These temperatures are low enough to give 
  a well-defined phenomenological rate when the reactive
  flux method is employed in the constant friction case,
but not so low that trajectories are needlessly slow
even when one obtains the rate by direct methods.

\subsection{Rate Formula and Numerical Methods}\label{ssec:numerical}

  The standard approach for calculating reaction rates, ``the reactive flux
  method,'' assumes stationarity.\cite{rmp90,chan78} 
  In establishing its validity, the rate formula needs to be checked
  by comparison with direct methods 
measuring the phenomenological rates between reactants and products.
In this section, a direct approach for obtaining
rates in the nonstationary cases of interest to this work is
reviewed and similarly validated.  
The results of this approach are subsequently used to 
motivate an averaged reactive flux formula appropriate for
the nonstationary case.

In the direct approach, one simply calculates the rate of population
transfer from the reactant population $n_{\rm a}$ to the
the product population $n_{\rm c}$.
The initial population
is assumed to be thermally distributed entirely at the reactant side.
  The latter assumption 
  is valid because the Boltzmann distribution is the steady state 
  solution of the system restricted to the reactant
  region (App.~\ref{sec:rflux}).
Assuming that a simple first-order master equation describes the rate
process (App.~\ref{sec:rflux}), 
the population in the reactant well can be solved 
directly as,
  \begin{equation}
  {{n_{\rm a}(t)-\bar{n}_{\rm a}}\over{n_{\rm a}({t_0}) -\bar{n}_{\rm a}}}
    =\exp\left(-\int_{t_0}^t dt' \lambda'(t)\right)\;,
  \end{equation}
  where the relaxation rate $\lambda(t)=k^++k^-$
is the sum of the forward ($k^+$) and reverse ($k^-$) rates, 
  $\bar{n}_{\rm a}$ is the population in the left 
  well at equilibrium.
  At equilibrium, for a symmetric potential, $n_{\rm a}(t)=n_{\rm c}(t)={N/2}$, where $N$ is the total population
of Brownian particles.
  In a nonequilibrium bath, such as is seen in the model
described in Sec.~\ref{sec:model}, that has oscillatory components
with a maximum recurrence time, then a phenomenological 
rate may still be obtained by averaging at sufficiently long times
compared to the maximum recurrence time.
In particular, 
\begin{subequations}\label{eq:20}\label{eq:intrate}%
\begin{eqnarray}
   \bar{\lambda}^{(1)} &\equiv& {1\over t-{t_0}}\int_{t_0}^t\lambda(t')dt'\\
  &=&-{1\over t-{t_0}} 
    \ln\left[{{n_{\rm a}(t)-{N/2}}
   \over{{N/2}}}\right] \;.
  \end{eqnarray}\end{subequations}%
The second equality was introduced by Pollak and Frishman\cite{frish93a}
as a construction that can lead to long time stability thereby
ensuring a substantial plateau time.\cite{chan78}
The instantaneous flux can be found using the differential expression
\cite{frish93a}:
\begin{subequations}\label{eq:intrate2}%
\begin{eqnarray}
   -\lambda(t) &=&{1\over{n_{\rm a}-n_{\rm c}}}{d\over{dt}}{(n_{\rm a}-n_{\rm c})} \\
          &=& {d\over{dt}}\ln{(n_{\rm a}-n_{\rm c})}\;.
\end{eqnarray}\end{subequations}%

  The numerical calculation of either of the direct rate
formulas requires the direct integration of a large number of trajectories
all initiated in the reactant region.
Consequently, it will only be accurate when the numerical integrator
is accurate for times that are sufficiently long to capture the
rate process.
This holds at the moderate temperatures (near $\beta V^{\ddagger}=10$)
explored in this work for which
  Eqs.~\ref{eq:intrate} and \ref{eq:intrate2} lead to the same result. 
The first method was used in all of the direct
calculations in this work because it tends to be more stable.

  The direct methods are time consuming and it is
  practically impossible to apply them
at low temperatures.
  As was mentioned at the beginning of this section,
the typical solution of this problem is the use of the 
reactive flux method.
It samples only those states that
traverse the dividing surface.
  For stationary systems, the
reactive flux is\cite{rmp90}
  \begin{eqnarray}
   k^+={{\langle\delta[q(0)]\dot{q}(0)\theta[q(t)]\rangle}\over{\langle\theta
          (q)\rangle}}\;,
\label{eq:rflux}
  \end{eqnarray}
where the characteristic equation
  $\theta[q(t)]$ for a trajectory is $1$ if
  the particle is in the right well at time $t$ and zero otherwise,
and the Dirac $\delta$-function ensures that
all the particles are initially located at the barrier (at $x=0$).
  The angle brackets represent the averaging over the thermal
  distribution of the initial conditions. 

\begin{table}
 \begin{tabular}{c r@{}l@{}l r@{}l@{}l r@{}l@{}l r@{}l@{}l}  
   & \multicolumn{12}{c}{$\omega$}\\ \cline{2-13}
  {Rates at $\gamma=10$}  
      &\multicolumn{3}{c}{0} &\multicolumn{3}{c}{0.1}
      &\multicolumn{3}{c}{1} &\multicolumn{3}{c}{10} \\ \hline
    integral method & 
      2&   &$\times10^{-6}$ 
     &1&.17&$\times10^{-5}$ 
     &1&.55&$\times10^{-5}$ 
     &8&.5 &$\times10^{-5}$ \\ 
    reactive flux &
      2&   &$\times10^{-6}$ 
     &2&   &$\times10^{-6}$ 
     &2&   &$\times10^{-6}$ 
     &2&.3 &$\times10^{-6}$ \\ \hline\hline
   & \multicolumn{12}{c}{$\omega$}\\ \cline{2-13}
  {Rates at $\gamma=1$}  
      &\multicolumn{3}{c}{0} &\multicolumn{3}{c}{0.1}
      &\multicolumn{3}{c}{1} &\multicolumn{3}{c}{10}\\ \hline
    integral method & 
      1&.7 &$\times10^{-5}$ 
     &2&.2 &$\times10^{-5}$ 
     &2&.9 &$\times10^{-5}$ 
     &3&   &$\times10^{-5}$ \\ 
    reactive flux &
      1&.6 &$\times10^{-5}$ 
     &1&.6 &$\times10^{-5}$ 
     &1&.6 &$\times10^{-5}$ 
     &2&.5 &$\times10^{-5}$ \\ \hline\hline
   & \multicolumn{12}{c}{$\omega$}\\ \cline{2-13}
  {Rates at $\gamma=0.05$}  
      &\multicolumn{3}{c}{0} &\multicolumn{3}{c}{0.1}
      &\multicolumn{3}{c}{1} &\multicolumn{3}{c}{10}\\ \hline
    integral method & 
      3&.3 &$\times10^{-5}$ 
     &1&.55&$\times10^{-5}$ 
     &1&.47&$\times10^{-5}$ 
     &2&.05&$\times10^{-5}$ \\ 
    reactive flux &
      3&   &$\times10^{-5}$ 
     &3&.15&$\times10^{-5}$ 
     &1&.76&$\times10^{-5}$ 
     &1&.9 &$\times10^{-5}$ \\ 
\end{tabular}
\caption{The integral method of Eq.~\protect\ref{eq:20} is compared
to the stationary reactive flux method of Eq.~\protect\ref{eq:rflux}
in calculating the activated rate across the double-well
potential in a rotating field of frequency $\omega$.
All the calculations are performed at the same bath temperature
such that $\beta V^{\ddagger}=10$,
and at three different values of $\gamma$ illustrative
of the low, intermediate and high friction limits.
Here and elsewhere, 
all values are reported in the dimensionless units 
of Eq.~\protect\ref{eq:dwpot}.
}\label{tab:rflux}
\end{table}

One might naively assume that the rate expression in Eq.~\ref{eq:rflux}
might still hold in the nonstationary case of time periodic friction,
Eq.~\ref{eq:tfric}.
The direct and reactive-flux rates at
  different frequencies and different friction constants
  are compared
  in Table \ref{tab:rflux}. 
  The two
  don't always agree and the difference can be as much as an order of magnitude.
  This result 
should not be surprising because of
  the nonstationarity of the problem.
However, correlation functions for this system do become stationary 
  when one averages over the period
  of the external perturbation. 
This suggests that Eq.~\ref{eq:rflux} should be further 
averaged over the initial
  time during a period of the external field,
yielding the average reactive flux rate,
  \begin{equation}
\label{eq:trflux}
  \bar\kappa(t)={\omega\over{2\pi}}\int_0^{2\pi\over{\omega}}d\tau{\langle
  {\delta
  [x(\tau)]\dot{x}(\tau)\theta[x(t+\tau)]\rangle}\over{\langle \theta\rangle}}\;.
  \end{equation}
(There is a formal proof in Appendix \ref{sec:rflux}.)
  A comparison between the direct rates and the average
  reactive flux rate is presented in Table \ref{tab:aveflux}.
\begin{table*}
\begin{tabular}{cr@{}l@{}lr@{}l@{}lr@{}l@{}lr@{}l@{}lr@{}l@{}l}
   & \multicolumn{15}{c}{$\gamma$}\\ \cline{2-16}
  $\omega$  
     & \multicolumn{3}{c}{.005}
     & \multicolumn{3}{c}{.05}
     & \multicolumn{3}{c}{.5}
     & \multicolumn{3}{c}{1}
     & \multicolumn{3}{c}{10} \\ \hline
   .1  & 3&   &$\times10^{-6}$ 
       & 1&.55&$\times 10^{-5}$ 
       & 2&.3 &$\times 10^{-5}$
       & 2&.1 &$\times 10^{-5}$ 
       & 1&.15&$\times 10^{-5}$ \\ 
       &(2&.78&$\times 10^{-6}$) 
       &(1&.51&$\times 10^{-5}$) 
       &(2&.3 &$\times 10^{-5}$)
       &(2&.1 &$\times 10^{-5}$)
       &(1&.18&$\times 10^{-5}$) \\ \hline
   .5  & 2&.8  &$\times 10^{-6}$  
       & 1&.43 &$\times 10^{-5}$ 
       & 2&.22 &$\times 10^{-5}$ 
       & 2&.1  &$\times 10^{-5}$ 
       & 1&.375&$\times 10^{-5}$ \\ 
       &(2&.62 &$\times 10^{-6}$) 
       &(1&.45 &$\times 10^{-5}$) 
       &(2&.23 &$\times 10^{-5}$) 
       &(2&.13 &$\times 10^{-5}$)
       &(1&.41 &$\times 10^{-5}$)\\ \hline
  1.0  & 3&   &$\times 10^{-6}$ 
       & 1&.45&$\times 10^{-5}$ 
       & 3&   &$\times 10^{-5}$
       & 2&.7 &$\times 10^{-5}$  
       & 1&.45&$\times 10^{-5}$ \\ 
       &(2&.7 &$\times 10^{-6}$) 
       &(1&.39&$\times 10^{-5}$) 
       &(3&.01&$\times 10^{-5}$) 
       &(2&.7 &$\times 10^{-5}$) 
       &(1&.47&$\times 10^{-5}$) \\ \hline
   5.0 & 3&   &$\times 10^{-6}$ 
       & 1&.9 &$\times 10^{-5}$ 
       & 3&.25&$\times 10^{-5}$ 
       & 2&.8 &$\times 10^{-5}$ 
       & 1&.2 &$\times 10^{-5}$ \\ 
       &(2&.7 &$\times 10^{-6}$) 
       &(1&.86&$\times 10^{-5}$) 
       &(3&.26&$\times 10^{-5}$) 
       &(2&.86&$\times 10^{-5}$) 
       &(1&.14&$\times 10^{-5}$)\\ \hline 
  10.0 & 3&   &$\times 10^{-6}$ 
       & 1&.92&$\times 10^{-5}$ 
       & 3&.25&$\times 10^{-5}$
       & 2&.7 &$\times 10^{-5}$  
       & 8&.6 &$\times 10^{-6}$ \\ 
       &(2&.7 &$\times 10^{-6}$) 
       &(1&.92&$\times 10^{-5}$) 
       &(3&.35&$\times 10^{-5}$)
       &(2&.78&$\times 10^{-5}$) 
       &(8&.87&$\times 10^{-6}$) \\ 
\end{tabular}
\caption{The integral method of Eq.~\protect\ref{eq:20} is compared
to the average reactive flux method of Eq.~\protect\ref{eq:trflux}
in calculating the activated rate across the double-well
potential in a rotating field of frequency $\omega$
and various friction constants $\gamma$.
The potential and inverse temperature ($\beta V^{\ddagger}=10$) 
are the same as in Table \protect\ref{tab:rflux}.
At each entry, the integral method result is written above
the more approximate average reactive flux result.
To aid the eye, the latter is also signaled by parentheses.
}\label{tab:aveflux}
\end{table*}
The numerics were performed at a temperature ($\beta V^{\ddagger}=10$)
that  
  is high enough to enable direct calculation
  of the rate within a few hours of CPU time on a current workstation.
  As can be seen from the table, there is very
  good agreement between the methods.
Equation \ref{eq:trflux} is the central result of this article,
and represents the fact that
{\it the reactive flux method is valid for the case of a time-dependent bath
   when a proper averaging is taken over the period of the external field}.
   This result is critical for the numerical calculation of rates because
   the direct approaches are cost prohibitive when
the temperature is much smaller then barrier height.
In this section the reactive flux method has been generalized to include 
   out-of-equilibrium systems 
  in cases in which an external force perturbs a bath 
  that is coupled to a reactive system. 
The resulting thermal flux is defined
only after averaging over the time period of the external perturbation.
 Using the non averaged rate expression would lead to undefined
  rates because the reactive system is so far out of equilibrium.
A detailed explanation can be found in App.~\ref{sec:rflux}.

\subsection{Analytical Approximations}\label{ssec:analytic}
  \subsubsection{Weak Friction}
  For the stationary problem, 
  Melnikov and Meshkov\cite{mm86} developed a perturbation
  technique to find the reactive flux at weak to moderate
  friction limit.\cite{pgh89} 
  The expansion parameter of the method is 
  the energy loss $\delta$ 
that
  a particle starting at the barrier 
experiences 
  while traversing the well.
Its value is
  \begin{equation}
   \label{eq:delta}
   \delta=\gamma\beta s\;,
  \end{equation}
where $s$ is the action of a frictionless particle starting with zero momentum at 
  the barrier and returning back to the top of the barrier  
after traversing a periodic orbit (the instanton), {\it i.e.},
\begin{equation}
  s=\int_{-\infty}^{\infty}p^2(t) dt=\int_{q(-\infty)}^{q(\infty)} pdq\;.
\end{equation}
  The resulting rate is
  \begin{equation}
  k=k_{\rm TST}\Upsilon\;,
\label{eq:StRate}
  \end{equation}
where $k_{\rm TST}={\omega\over{2\pi}}e^{-\beta V^{\ddagger}}$ is the transition state 
  theory rate ($\omega$ is the frequency at the bottom of the reactant well and
  $V^{\ddagger}$ is the barrier height) and the depopulation factor $\Upsilon$ is
  \begin{eqnarray}
   \label{eq:tra}
   \Upsilon(\delta)=\exp\left\{{1\over{2\pi}}\int_{-\infty}^{\infty}\ln
   \left[1- e^{ \delta(\lambda^2+1/4) }
   \right]{1\over{\lambda^2+1/4}
   }d\lambda\right\}.
  \end{eqnarray}

The nonstationary analytic rate expression can now be obtained by
analogy to the formulation of the average reactive flux rate 
in which the rate is averaged 
over the period of driving term.
In particular, the energy loss,
  \begin{equation}
   \delta(\tau)=\gamma\beta\int_{-\infty}^{\infty}\psi(t+\tau)^2p^2(t)dt\;,
\label{eq:dtau}
  \end{equation}
is now obtained as a function
of the possible initial configurations of the driving term
which are, in turn, parameterized by the time lag $\tau$  
relative to the start of an oscillation 
in the friction of Eq.~\ref{eq:tfric}.
Trajectories in one dimension can be calculated up to a quadrature
directly from energy conservation,
\begin{equation}
q(t) = q(t_0) + 2\int_{t_0}^{t} \!dt\,\sqrt{E-V(q)}\;.
\end{equation}
In the case of the double-well potential defined by
Eq.~\ref{eq:quar}, the instanton at $E=V^{\ddagger}$
---{\it viz.}~the periodic orbit on the upside-down potential---
can be obtained analytically.
The results for time, 
  \begin{equation}
  t(q)=-{1\over 2}\ln\left({\sqrt{2}+\sqrt{2-q^2}\over q}\right)\;,
  \end{equation}
   and momentum,
  \begin{eqnarray}
  \label{eq:pt}
  p(q)&=&\sqrt{2}q\sqrt{2-q^2}\;,
 \end{eqnarray}
as a function of $q$ follow readily.
By substitution into Eq.~\ref{eq:dtau},
the energy loss parameter is obtained directly with respect to the
time lag $\tau$ relative to the start of an oscillation
in the friction of Eq.~\ref{eq:tfric}, {\it i.e.},
 \begin{eqnarray}
  \delta(\tau)&=&
     2\gamma\beta\int_0^{\sqrt{2}}\!dq\,
         \Biggl\{a
   \Biggr.  \nonumber\\ 
&&+ \left.
      \cos\left[-{\omega\over 2}\ln
             \left({\sqrt{2}+\sqrt{2-q^2}\over 2}
              \right)+\omega\tau
           \right]
      \right\}
\nonumber \\ 
    &&\times \sqrt{2}q\sqrt{2-q^2}
\;.
 \end{eqnarray}
The nonstationary rate formula for a time-periodic driving friction
can thus be written as the product of the TST rate and 
a generalized depopulation factor averaged over $\tau$,
\begin{equation}
\bar\Upsilon[\delta]\equiv\int_{0}^{1}\!d\tau\,
   \Upsilon\bigl(\delta(\tau)\bigr)
\;,
\label{eq:NstRate}
\end{equation}
in analogy with Eq.~\ref{eq:StRate}.

 The validity of the analytical result of Eq.~\ref{eq:NstRate}
 for the rate can be checked in the
 low friction regime in which $\Upsilon(\delta)\approx\delta$.
 Taking the average over a period yields the result,
\begin{eqnarray}
 \bar{\Upsilon}[\delta] &\approx&  
  {\omega\over{2\pi}}\int_0^{2\pi/\omega}
   \delta(\tau)d\tau 
  \nonumber\\
   &=&{8\over 3} \left(a^2+{ 1\over 2}\right)\;.
\label{eq:32}
\end{eqnarray}
This result is in good agreement with
the averaged reactive flux rate formula of Eq.~\ref{eq:trflux},
as shown in Table \ref{tab:transmission}
at a low friction  value ($\gamma_0=.005$),  
$\beta V^{\ddagger}=10$, and various frequencies. 
Even within this weak friction regime, as the friction increases, the
approximation leading to Eq.~\ref{eq:32} will break down.
The direct evaluation of Eq.~\ref{eq:NstRate} corrects this 
error, and also leads the rate to depend on the frequency of the
driven friction.

\begin{table}
 \begin{tabular}{cr@{}l@{}lr@{}l@{}lr@{}l@{}l}
  &\multicolumn{9}{c}{$\omega$}\\ \cline{2-10}
  {Rates at $\gamma=0.005$}  
      & \multicolumn{3}{c}{0.1} & \multicolumn{3}{c}{1}
          & \multicolumn{3}{c}{10} \\ \hline
  MM (Eq.~\protect\ref{eq:32}) 
       & 8&.54&$\times10^{-2}$ 
       & 8&.54&$\times10^{-2}$ 
       & 8&.54&$\times10^{-2}$\\ 
  $\bar k(t)$ (Eq.~\protect\ref{eq:trflux}) 
       & 7&.5&$\times10^{-2}$ 
       & 7&  &$\times10^{-2}$ 
       & 7&.5&$\times10^{-2}$\\ 
 \end{tabular} 
  \caption{The transmission coefficients for the escape rate $k$
across a quartic potential at $\beta V^{\ddagger}=10$ and
$\gamma=.005$ obtained 
using the average reactive flux method of Eq.~\protect\ref{eq:trflux}
with the analytical Melnikov-Meshkov expression \protect\ref{eq:32}
  for $\bar{\delta}$.
An ensemble of 100,000 trajectories has been propagated in each of 
the reactive flux calculations.
}\label{tab:transmission}
\end{table}

  \subsubsection{Strong Friction}

  The reaction rate in the overdamped regime of the 
LE 
  is well known.\cite{rmp90}
 The central idea is that the motion in phase space is 
 strongly diffusive in this regime.
  The rate 
is consequently dominated by
  the dynamics close to 
  the barrier.  
  At the vicinity of the barrier top, the potential can be 
  approximated 
by an inverted parabolic potential 
  and the LE at the barrier can be written as
  \begin{eqnarray}
  \label{eq:cf}
  \ddot{q}=-\omega_{\rm b}^2 q-\gamma \dot{q} + \xi(t)\;,
  \end{eqnarray}
where the fixed friction $\gamma$ and stochastic force $\xi(t)$ satisfy
the regular fluctuation dissipation relation (Eq.~\ref{eq:fdr}) 
  and $i\omega_{\rm b}$ is the imaginary frequency at the barrier.
  It was shown that the reaction rate for this case is\cite{grot80,pollak86b}
  \begin{eqnarray}
  \label{eq:tot}
   k={\lambda_{\rm b}\over{\omega_{\rm b}}}{\omega_0\over{2\pi}}\exp{-\beta V^{\ddagger}}\;,
  \end{eqnarray}
where $\omega_0$ is the frequency of the reactant well, and
$i \lambda_{\rm b}$ is the imaginary eigenvalue of the 
  homogeneous part of Eq.\ref{eq:cf}.
  The latter is related to the exponential 
  divergence in the trajectories near the barrier,
  \begin{eqnarray}
  \label{eq:sle}
   q(t)\propto e^{\lambda_{\rm b} t}.
  \end{eqnarray}

At strong friction in the nonstationary problem, 
  the reaction rate expression is also dominated
  by the trajectories in the barrier region.
Equation \ref{eq:tot} can still be 
  used for the rates, though now $\lambda(t)$ is the time-dependent 
eigenvalue of the homogeneous part,
  \begin{eqnarray}
  \label{eq:th}
   \ddot{q} + \gamma\phi(t)\dot{q} -\omega_{\rm b}^2 q=0\;,
  \end{eqnarray} 
  of the nonstationary stochastic equation of motion.
  The solution of this equation is not trivial.
  A possible way to solve the problem is found in Ref.~\onlinecite{RH97}.
It is easier to extract the eigenvalue
  numerically from the exponential divergence of trajectories
  starting near the barrier top,
  \begin{eqnarray}
   \label{eq:sile}
   q(t)\propto e^{\int^t\lambda_{\rm b}(t')dt'}\;.
  \end{eqnarray} 
  The periodicity of the time dependent coefficient in Eq.~\ref{eq:th} 
leads also to a
  periodicity in $\lambda_{\rm b}(t)$. 
  If $\lambda_{\rm b}$ is the time average
   of the time-dependent eigenvalue of Eq.~\ref{eq:th}
   over a period, 
  then for $t$ much larger than the period,
   Eq.~\ref{eq:sile} is analogous Eq.~\ref{eq:sle}
   with $\bar{\lambda}_{\rm b}$ in the exponent.
Replacement of $\lambda_{\rm b}$ by $\bar\lambda_{\rm b}$ in the rate expression
(Eq.~\ref{eq:tot})
provides good agreement with the averaged reactive flux rates
as shown in the high friction columns of Table \ref{tab:eq38}.

  \subsubsection{Weak to High Friction}

  The results of the two previous subsections have motivated
the redefinition of the components of the rate formula in the
low and high friction limits of the nonstationary time-periodic problem.
Retaining these assignments in the stationary turnover rate formula\cite{pgh89}
suggests the nonstationary turnover rate,
  \begin{equation}
  \bar{k}={\bar{\lambda}_{\rm b}\over{\omega_{\rm b}}}{\omega_0\over{2\pi}}
   { {\bigl({\bar\Upsilon[\delta]}\bigr)^2}\over{\bar{\Upsilon}[2\delta]} 
   }
      \exp{-\beta V^{\ddagger}}\;.
\label{eq:38}
  \end{equation}
\begin{figure}
\centerline{\psfig{figure=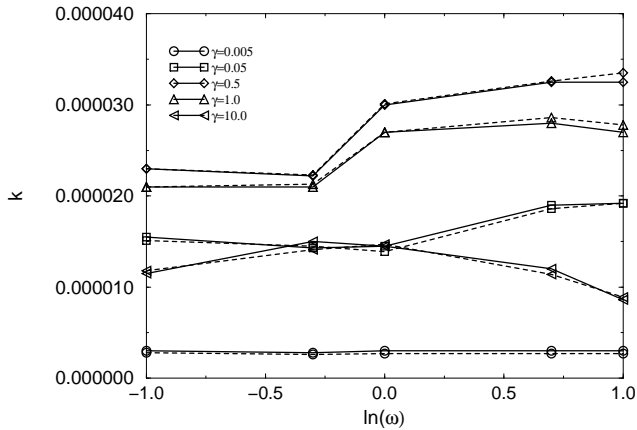,height=2.5in,angle=-90}}
 \caption{The activated escape rates $k$ of particles
in a quartic potential and solvated by an anisotropic
time-dependent liquid is obtained as a function
of the driving frequency $\omega$ using two numerical methods
described in this work.
The numerical direct rate of Eq.~\protect\ref{eq:20} is
shown by dashed lines,
and the averaged reactive flux rate of Eq.~\protect\ref{eq:trflux} 
is shown by solid lines. 
In the former, an ensemble of 250,000 initial conditions were
used to achieve convergence, and the corresponding 
numerical values 
   are summarized in Table \protect\ref{tab:transmission}.
In the latter, the average was performed
  over an ensemble of 100,000 trajectories, yielding the results
  in a wall-clock time that was an order of magnitude faster
than that for the direct rate calculations.
In all cases, the inverse temperature $\beta V^{\ddagger}$ is 10.
}
\label{fig:rate}
\end{figure}
\begin{figure}
\centerline{\psfig{figure=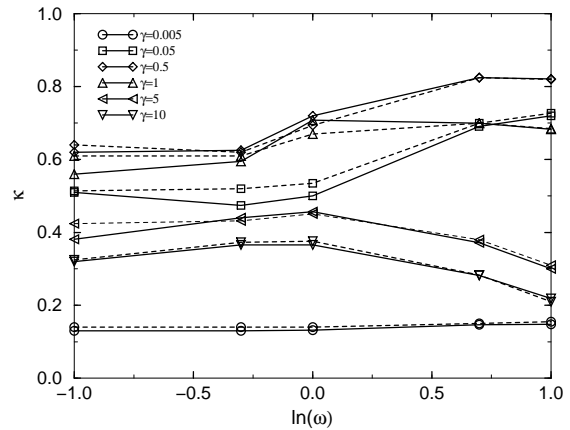,height=2.5in,angle=-90}}
\caption{The activated escape rates $k$ of particles
in a quartic potential and solvated by an anisotropic
time-dependent liquid is obtained as a function
of the driving frequency $\omega$ comparing the reactive flux
approach to an analytical result.
(The latter is expected to be accurate  
here---and not in Fig.~\protect\ref{fig:rate} 
or Table \protect\ref{tab:aveflux}---because the inverse temperature
has been increased to 20.)
%
%
The solid line corresponds to the numerical result calculated 
using the averaged reactive flux rate
  method of Eq.~\protect\ref{eq:trflux} and the dashed line 
corresponds to the turnover formula 
in Eq.~\protect\ref{eq:38}.
  The numerical calculations were performed by averaging
over an ensemble of 250,000 trajectories.}
\label{fig:kappa}
\end{figure}
\begin{table}
\vspace{6pt}
 \begin{tabular}{cl@{\,}ll@{\,}ll@{\,}ll@{\,}ll@{\,}ll@{\,}l}
   & \multicolumn{12}{c}{$\gamma$}\\ \cline{2-13}
  $\omega$  
     & \multicolumn{2}{c}{.005}
     & \multicolumn{2}{c}{.05}
     & \multicolumn{2}{c}{.5}
     & \multicolumn{2}{c}{1}
     & \multicolumn{2}{c}{5}
     & \multicolumn{2}{c}{10} \\ \hline
   .1  & (&.14)  & (&.514)  & (&.64)  & (&.61)  & (&.424) & (&.325) \\
       &  &.13   &  &.51    &  &.62   &  &.56   &  &.381  &  &.32 \\ \hline
   .5  & (&.14)  & (&.52)   & (&.62)  & (&.61)  & (&.432) & (&.373) \\
       &  &.13   &  &.474   &  &.625  &  &.595  &  &.44   &  &.366\\ \hline
  1.0  & (&.14)  & (&.535)  & (&.694) & (&.67)  & (&.451) & (&.376) \\
       &  &.132  &  &.5     &  &.72   &  &.708  &  &.457  &  &.366\\ \hline
  5.0  & (&.15)  & (&.7)    & (&.825) & (&.7)   & (&.38)  & (&.283) \\
       &  &.147  &  &.691   &  &.825  &  &.7    &  &.372  &  &.282\\ \hline
 10.0  & (&.155) & (&.727)  & (&.82)  & (&.685) & (&.31)  & (&.21)  \\
       &  &.148  &  &.720   &  &.821  &  &.683  &  &.3    &  &.219
\end{tabular} 
  \caption{The average reactive flux method of Eq.~\protect\ref{eq:trflux}
is compared to the analytic approximation of Eq.~\protect\ref{eq:38} 
in calculating the activated rate across the quartic
potential in a rotating field of frequency $\omega$
and various friction constants $\gamma$.
The inverse temperature ($\beta V^{\ddagger}=20$) is higher
in contrast to the previous tables.
At each entry, the more approximate average reactive flux result
is written above the analytic result.
To aid the eye, the former is also signaled by parentheses.
}\label{tab:eq38}
\end{table}
The prefactors from the nonstationary turnover rate are compared
to those from the averaged reactive flux rate at the
inverse temperature 
  $\beta=20$ in Table \ref{tab:eq38} 
  and in Fig.~\ref{fig:kappa}.
  As can be seen, there is a very good agreement between the numerical and
  analytic results at the very weak and strong friction limits. 
  Therein the results differ by no more than 5\% 
  throughout the frequency range; an error margin smaller 
  than the error bars in the numerical calculations. 
  At moderate friction
  and low frequencies, however, the differences 
---on the order of 10\%--- cannot be explained
  by error in the numerical calculations alone, 
and may be significant.
Corrections or improvements in the approximations
leading to the connection formula of Eq.~\ref{eq:38} are
also of interest, but not pursued further in this work.
  Recall that the turnover escape rate expression for the LE with constant
  friction was obtained through the solution of the
  equivalent Hamiltonian formalism.\cite{pgh89}
  A similar approach for the solution of a
  the Hamiltonian equivalent\cite{hern99e} of
  the stochastic time-dependent bath problem may
  lead to a fruitful solution.
However, even in the 
  constant friction case, the turnover formula can give 
rise to small systematic error. 
With these reservations, the approximate rate formula
  can be used to obtain time-dependent escape rates.
  It is clear from the results that there is a frequency effect on the
  reaction rates.
  For the specific example studied here,
  the effect can modify the reaction rate

 \section{Discussion and Conclusions}
\label{sec:conc}

In this work, several techniques for obtaining the dynamics of interacting
  Brownian particles that are coupled to a time dependent thermal bath
  have been discussed.
  Two models, one of dynamics in lyotropic
  liquids and one for dynamics in pure nematic liquid under a periodic external field has been brought as  examples of such systems. 
  The models include a new mechanism for stochastic dynamics in which 
  an external force is used to drive the thermal bath.
  There is no net injection of energy to the 
  Brownian particles in the bath due to the driving force, hence
  they keep their equilibrium properties. Yet observables
  such as reaction and diffusion rates are modified. 
  The existence of a steady state that retains the equilibrium 
  enables one to express
  out-of-equilibrium observables with respect to averaging over the equilibrium.
  This is the Onsager regression hypothesis 
(Appendix \ref{sec:rflux}). 
  We used this to 
  extend known methods for calculating the reaction rates in the constant 
  friction to nonstationary baths.
  Extensive computation effort was used to illustrate the diffusive
  and reactive rates for an effective Brownian particle in the
  naive anisotropic liquid bath model with rotating magnetic field. 
However, the numerical and analytical tools that have been
  modified and developed in this work are
  appropriate for any model with time-dependent friction. 
 
The construction introduces new
  control parameters into the problem; 
namely,
  the external force amplitude and
  frequency. 
  We concentrated on the latter and exhibited the frequency dependence of 
  diffusion and reaction rates in the naive model. 
  This dependence is not linear and
  changes dramatically with the friction strength. 
  The enhancements in the 
  reaction rate and the diffusion coefficient are not the same, 
{\it i.e.}, the maximum
  in the diffusion rate as a function of the external frequency is not the same
  as the maximum in the rate.
  This nonlinear behavior could be used to enhance reaction diffusion
  processes, such as cluster nucleation, by up to a few order of magnitude.

  In the extension of the naive model to more realistic nematic liquids, the 
  cooperative effects of the liquid cannot be omitted. 
  There are
  phenomenological difficulties in defining friction and the fluctuation 
  dissipation relation in liquid crystals.
  To the best of
  our knowledge such a theory is still not fully developed.
The development of such a theory based on microscopic assumptions
is an extremely challenging problem.
  Boundary effects and elastic forces will create dynamical micro-domains
characterized by differing uniform directors 
  in a real nematic under rotating magnetic field.
  The theory for dynamics in nematics will have to deal also
  with the spatial inhomogeneities.
  These are among the challenges to future work in trying to better
  understand the diffusive dynamics in lyotropic liquids.
  
\section*{ACKNOWLEDGEMENTS}
We are grateful to Prof. Rina Tannenbaum for stimulating discussions and
helpful suggestions.
This work has been partially supported by 
the National Science Foundation under Grant,
Nos.~97-03372 and 02-13223.
RH is a Goizueta Foundation Junior Professor.
The Center for Computational Science and Technology
is supported through a Shared University Research (SUR)
grant from IBM and Georgia Tech.

\appendix

  \section{Fourth-order integrator for the LE with periodic friction }
\label{sec:integrator}

A high-order integrator was developed for the regular LE or GLE in 
  Ref.~\onlinecite{art2}.
   A modified algorithm for time- and space-dependent friction was developed
  for the explicit GLE with exponential memory kernel in the friction
  \cite{herher01}.
  This appendix introduces the numerical scheme 
necessary for solving a
  time dependent
  stochastic equation equation of motion of the form of Eq.~\ref{eq:tlan}.

  A finite difference scheme is used to propagate the solution over a small
  time step. 
 At each iteration, the propagator is expanded to fourth order with
  respect to the time step using a strong Taylor scheme.\cite{kloe92}
  The resulting integrator can be decomposed into two uncoupled terms:
\begin{subequations}\begin{eqnarray}
    q(t+h)&=&q_{\rm det}(p,q,t)+q_{\rm ran}(p,q,t)\\ 
    p(t+h)&=&p_{\rm det}(p,q,t)+p_{\rm ran}(p,q,t)\;.  
  \end{eqnarray}\end{subequations}%
  The deterministic terms that are collected within $
  q_{\rm det}$ and $p_{\rm det}$
  are those that remain in a fourth-order Taylor expansion of 
  the deterministic equation of
  motion after removing any term that includes a stochastic variable. 
  The deterministic propagator can be calculated numerically with any fourth-order
  deterministic scheme; the fourth order Runge Kutta method
  was chosen for this work.\cite{numrec}

  The random propagator 
(giving rise to the stochastic contribution, $q_{\rm ran}$ and $p_{\rm ran}$)
  includes all terms up to fourth-order that include stochastic variables. 
  For the present case, the random integrator for the space and momenta 
leads to the contributions,
 \begin{eqnarray}
  q_{ran}&=&a\left\{\left[\psi+\dot{\psi}+{1\over 2}\ddot{\psi}h^2\right]
          Z_2
      \right.\nonumber\\
   &&+\left.
    \left[-\gamma\psi^3-2\dot{\psi}-(2\ddot{\psi}+3\gamma
          \psi^2\dot{\psi})h^2\right]Z_3\right. \nonumber\\
    &&+\left.\left[-V''\psi+\gamma^2\psi^5+2\ddot{\psi}+
          7\gamma\psi^2\dot{\psi}\right]Z_4\right\} \\ 
  p_{ran}&=&a\left\{\left[\psi+\dot{\psi}+{1\over 2}\ddot{\psi}h^2+
           {1\over 6}\psi^{(3)}h^3\right]Z_1 \right.\nonumber\\ 
         &+&\left[-\gamma\psi^3-\dot{\psi}-
             (\ddot{\psi}+3\gamma\psi^2\dot{\psi})h
    \right.\nonumber\\&&-\left.
             (-{1\over 2}\psi^{(3)}+\gamma(3\psi\dot{\psi}^2+{3\over 2}
             \psi^2\ddot{\psi}))h^2\right]Z_2 \nonumber\\ 
         &+&\left[-V^{''}\psi+\gamma^2\psi^5+\ddot{\psi}+4\gamma
             \psi^2\dot{\psi} \right.\nonumber\\
          &&+\left.(p^0 V^{'''}\psi+5\gamma^2\psi^4\dot
             {\psi}+4\gamma(2\psi\dot{\psi}^2+\psi^2\ddot{\psi})
    \right.\nonumber\\&&-\left.
             V^{''}\dot{\psi}+\psi^{(3)})h\right]Z_3 \nonumber \\ 
         &+&\left[2\gamma V^{''}\psi^3-\gamma^3\psi^7+
             p_0 V^{'''}\psi+
             3V^{''}\dot{\psi}-\psi^{(3)}
    \right.\nonumber\\&&-\left.\left.
             9\gamma^2\psi^4\dot{\psi}-
             4\gamma(2\psi\dot{\psi}^2+\psi^2\ddot{\psi})\right]Z_4
             \right\}\;,
 \end{eqnarray}
where $\psi(t)$ is the coefficient in Eq.~\ref{eq:tfric} dictating the
time-dependence in the friction, and
     the Gaussian random variables $\{Z_i\}$
are correlated according to the moments specified  
        in Ref.~\onlinecite{art2}.
The symbol, $\psi^{(3)}$, with the parenthesized superscript,
denotes the third-order derivative of $\psi$ with respect to time.
     There is frequent repetition of terms differing by no more
than a ratio of constant coefficients, and this allows the integrator
to be computed very economically despite the seemingly large number
of terms contained above.
In fact, the most time-consuming part of the scheme is 
the calculation of the random numbers.
Although the fourth-order integrator has been expanded in terms of
the stochastic variables, the neglected higher-order terms have
coefficients that depend on the friction and frequencies to high order.
Consequently, the algorithm loses its efficiency at high
       friction or in the high frequency domain. 
For a given problem, a comparison of the fourth-order algorithm 
to lower-order algorithms such as the velocity
       Verlet algorithm\cite{allen87}
is advisable to ensure that the requisite accuracy is achieved.
Though not presented explicitly here,
such comparisons have been performed 
with favorable agreement for the models of this work.
The fourth-order integrator is accurate, and equally importantly,
provides a substantial savings in computational
time.
       
  \section{Reactive flux method for systems in nonstationary environment}
\label{sec:rflux}

  The reactive flux formalism
is an
  efficient numerical tool 
for calculating the
  escape rates of Brownian particle from a metastable well 
at
  low temperatures. The derivation of the reactive flux method is well
  formulated.\cite{rmp90}
In this appendix, the derivation is recapitulated in oder to emphasize
the new considerations that emerge because of nonstationarity.
  The corner stone for any rate calculation in a nonequilibrium
  system is 
 ``the Onsager regression hypothesis.''\cite{chandler} 
   This hypothesis asserts that
  ``the relaxation of macroscopic non-equilibrium disturbances is governed
   by the same laws as the regression of spontaneous microscopic
   fluctuations in an equilibrium system.''\cite{chandler}
   The two basic assumptions of the regression hypothesis are the existence of
   an equilibrium and 
a considerable (large) separation of time scales between the motion of the
subsystem and that of the overall relaxation of the system.

   The proof of the existence of an equilibrium in a 
system 
   is more readily obtained through an analysis of the 
   probability distribution $W(q,p;t)$
of the Brownian particles rather than the explicit trajectories
calculated using the LE. 
  The equation of motion for the distribution is the
Fokker-Planck equation,\cite{risken89}
 \begin{eqnarray}
 {\partial\over{\partial t}}W(q,p;t)
  &=&\left\{-{\partial\over{\partial q}}p-
  {\partial\over{\partial p}}\left[-V'(q)-\gamma_0\psi^2(t)p\right]
    \right.\nonumber\\&&-\left.
  +{\partial^2\over{\partial{p}^2}}{\gamma\over{\beta}}\psi^2(t)\right\}W(q,p;t)
\;,
 \end{eqnarray}
which corresponds to the nonstationary LE in Eq.~\ref{eq:tlan}.
 A direct substitution of the Boltzmann distribution confirms that it
 is indeed a solution of this 
 equation. 
 From this one can deduce that the Boltzmann distribution is the equilibrium
 distribution for systems with time-dependent friction.
 It might seems strange that equilibrium doesn't change 
when the system is driven by a
 time-dependent forcing friction. 
 The key point is that the force is coupled only to the bath.
 The bath is taken to be infinite dimensional 
 and absorbed the energy extract
 by the force. 
 This, of course, does not mean that the system dynamics is not
 modified.  On the contrary, as has been shown in the article,
 the diffusivity and
 reaction rates are modified by the periodicity of the externally
 driven friction. 
The time scales relevant to the second assumption are
 the period $\tau_{\rm f}$
 of the external force,  the transient time $\tau_{\rm s}$ of the 
 system, the typical escape time $\tau_{\rm e}$ of a thermal particle
 and the observation time $\tau $.
By construction of the naive model and the choice of its
parameterization, these times follow the simple
inequality:  $(\tau_{\rm f}\ \mbox{or}\ \tau_{\rm s})\ll\tau\ll\tau_{\rm e}$. 
As such, the system necessarily satisfies both of the assumptions 
needed to apply the Onsager hypothesis.

The rate constants may be obtained from a first order master
equation representing the population dynamics,
\begin{subequations}\label{eq:trate}\begin{eqnarray}
 \dot n_{\rm a}(t)=-k^+(t)n_{\rm a}(t)+k^-(t)n_{\rm c}(t) \\ 
 \dot n_{\rm c}(t)=k^+(t)n_{\rm a}(t)-k^-(t)n_{\rm c}(t)\;,
 \end{eqnarray}\end{subequations}%
where $n_{\rm a}$ ($n_{\rm c}$) is the population of the left (right) well of the
 quartic potential in Eq.~\ref{eq:quar}. 
The nonstandard feature
 is that the forward (backward) coefficient rate $k^{\pm}(t)$
 is time dependent. 
As described in the text, 
 they do have the same periodicity as the time-dependent friction
The master equation
in Eq.~\ref{eq:trate} can be integrated in the
usual manner to obtain the result, 
 \begin{equation}
  \label{eq:expra}
  {{\Delta n_{\rm a}(t)}\over{\Delta n_{\rm a}(t_0)}}=e^{-\int_{t_0}^t\lambda(t')dt'}\;,
 \end{equation}
where $\Delta n_{\rm a}(t)$ is the fluctuation of the momentary
 population of the left well relative to the equilibrium 
 population of the well, and $\lambda\equiv (k^+ + k^-)$ as in the text.
 The Onsager regression hypothesis enables the connection
  to the equilibrium averaged expression,
 \begin{equation}
  {{\langle\delta\theta[q(t_0)]\delta\theta[q(t)]\rangle}\over{
    \langle\delta\theta[q(t_0)]^2\rangle}}=e^{\int_{t_0}^t-\lambda(t')dt'}\;.
 \end{equation}
 Taking the time derivative of both sides leads to
\begin{subequations}\label{eq:derrate}\begin{eqnarray}
   {{\langle\delta\theta[q(t_0)]\dot\delta\theta[q(t)]}\over{
    \langle\delta\theta[q(t_0)]^2\rangle}}
  &=& -\lambda(t)e^{-\int_{t_0}^t\lambda(t')dt'}\\
   &\approx& -\lambda(t)\;.
 \end{eqnarray}\end{subequations}
  The last equality is a result of the large time scale separation.
  So far there is no real difference from the standard  derivation. 
In the usual derivation, the assumption of stationarity 
would now permit the modification of the numerator into the form
of a flux correlation function. 
  This cannot be performed in the present time-dependent case.
However, stationarity can be
  regained 
in this system
  by averaging over the period of the time-dependent friction. 
In practice, this can be achieved by initiating the subsystem at various
time shifts $\tau$ relative to some arbitrary time origin in the time-dependent
friction.
  An integration over the period of \ref{eq:derrate} leads to the averaged form
  of the reactive flux,
%
\begin{subequations}\label{eq:ar}\begin{eqnarray}   
  {\omega\over{2\pi}}\int_0^{2\pi\over{\omega}} &&d\tau
  {\langle\delta q[\tau][\dot{q}(\tau)]
  \theta[q(t+\tau)]\rangle\over{\langle\delta\theta[q(\tau)]^2\rangle}}
\nonumber\\
&=&
{\omega\over{2\pi}}\int_0^{2\pi\over{\omega}}d\tau
  {\langle\delta\theta[q(\tau)]
  \dot{\theta}[q(t+\tau)]\rangle\over{\langle\delta\theta[q(\tau)]^2\rangle}}\\
&\approx& {\omega\over{2\pi}}\int_0^{2\pi\over{\omega}}d\tau \lambda(\tau) \\
&\equiv& -\bar{\lambda}
 \end{eqnarray}\end{subequations}%
 The bar in the definition of $\bar\lambda$
 indicates the average over time period of the time-dependent friction.
 Equation \ref{eq:ar} is the averaged reactive flux rate for the 
 nonstationary system with periodic friction.
 The above algebra also justifies the extension of the Melnikov-Meshkov
  theory to the time
 dependent friction case averaged over a period of the external perturbation
described in Sec.~\ref{ssec:analytic}.

\bibliography{eli1}

\begin{thebibliography}{10}

\bibitem{risken89}
H. Risken, {\em The Fokker-Planck Equation} (Springer--Verlag, New York, 1989).

\bibitem{rmp90}
P. H{\"a}nggi, P. Talkner, and M. Borkovec, Rev. Mod. Phys. {\bf 62},  251
  (1990), and references therein.

\bibitem{yang}
H. Yang, Z. Zhuo, X. Wu, and X. Tang, Phys. Lett A {\bf 203},  157  (1995).

\bibitem{widata}
H.~W.~Y. Hsia, N. Fang, and X. Lee, Phys. Lett A {\bf 215},  326  (1996).

\bibitem{tucker01}
A.~N. Drozdov and S.~C. Tucker, J. Phys. Chem. B {\bf 105},  6675  (2001).

\bibitem{tucker01b}
A.~N. Drozdov and S.~C. Tucker, J. Chem. Phys. {\bf 114},  4912  (2001).

\bibitem{harris02}
K.~R. Harris, J. Chem. Phys. {\bf 116},  6379  (2002).

\bibitem{tucker02R}
A.~N. Drozdov and S.~C. Tucker, J. Chem. Phys. {\bf 116},  6381  (2002).

\bibitem{hern99a}
R. Hernandez and F.~L. Somer, J. Phys. Chem. B {\bf 103},  1064  (1999).

\bibitem{deggen93}
P.~G. deGennes and J. Prost, {\em The Physics of Liquid Crystals} (Clarendon
  Press, Oxford, 1993).

\bibitem{diogo}
A.~C. Diogo, Mol. Cryst. Lid. Cryst {\bf 100},  153  (1983).

\bibitem{ruhter}
R.~W. Ruhwandl and E.~M. Terentjev, Phys. Rev. E {\bf 54},  5204  (1996).

\bibitem{herher01}
E. Hershkovitz and R. Hernandez, J. Phys. Chem. A {\bf 105},  2687  (2001).

\bibitem{chrpat}
J. Christoph, P. Gabriel, and P. Davidson, Adv. Mater. {\bf 12},  9  (2000).

\bibitem{sramey89}
G. Srajer, S. Fraden, and R. Meyer, Phys. Rev. A {\bf 39},  4828  (1989).

\bibitem{brochard74}
F. Brochard, J. Phys. (Paris) Lett. {\bf 35},  L19  (1974).

\bibitem{meyer93}
K.~B. Migler and R. Meyer, Phys. Rev. E {\bf 48},  1218  (1993).

\bibitem{roman89}
V. Roman and E. Terent'ev, Colloid J. USSR {\bf 51},  435  (1989).

\bibitem{mm86}
V.~I. Mel'nikov and S.~V. Meshkov, J. Chem. Phys. {\bf 85},  1018  (1986).

\bibitem{pgh89}
E. Pollak, H. Grabert, and P. H{\"a}nggi, J. Chem. Phys. {\bf 91},  4073
  (1989).

\bibitem{lang69}
J.~S. Langer, Ann. Phys. (NY) {\bf 54},  258  (1969).

\bibitem{schuss82}
B.~J. Matkowski, Z. Schuss, and E. Ben-Jacob, SIAM (Soc. Ind. Appl. Math.) J.
  Appl. Math. {\bf 42},  835  (1982).

\bibitem{schuss83}
B.~J. Matkowski, Z. Schuss, and C. Tier, SIAM (Soc. Ind. Appl. Math.) J. Appl.
  Math. {\bf 43},  673  (1983).

\bibitem{berne84}
B. Berne, Chem. Phys. Lett. {\bf 107},  131  (1984).

\bibitem{bork85}
M. Borkovec and B. Berne, J. Chem. Phys. {\bf 82},  794  (1985).

\bibitem{nitzan88}
A. Nitzan, Adv. Chem. Phys. {\bf 70},  489  (1988).

\bibitem{herpol97}
E. Hershkovitz and E. Pollak, J. Chem. Phys. {\bf 106},  7678  (1997).

\bibitem{bsv81}
R. Benzi, A. Sutera, and A. Vulpiani, J. Phys. A {\bf 14},  L453  (1981).

\bibitem{march98}
L. Gammaitoni, P. H{\"a}nggi, P. Jung, and F. Marchesoni, Rev. Mod. Phys. {\bf
  70},  223  (1998).

\bibitem{DG92}
C.~R. Doering and J.~C. Gadoua, Phys. Rev. Lett. {\bf 69},  2318  (1992).

\bibitem{RH97}
P. Reimann and P. H{\"a}nggi,  in {\em Stochastic Dynamics}, Vol.~484 of {\em
  Lecture Notes in Physics}, edited by L. Schimansky-Geier and T. Poschel
  (Springer, Berlin, 1997), pp.\ 127--139.

\bibitem{manman00}
J. Lehmann, P. Reimann, and P. H{\"a}nggi, Phys. Rev. Lett. {\bf 84},  1639
  (2000).

\bibitem{mag93}
M.~O. Magnasco, Phys. Rev. Lett. {\bf 71},  1477  (1993).

\bibitem{Reim02}
P. Reimann, Phys. Rep. {\bf 361},  57  (2002).

\bibitem{Reim03}
A. Engel, H.~W. Mueller, P. Reimann, and A. Jung, Phys. Rev. Lett. {\bf 91},
  060602  (2003).

\bibitem{grot80}
R.~F. Grote and J.~T. Hynes, J. Chem. Phys. {\bf 73},  2715  (1980).

\bibitem{hang82}
P. H{\"a}nggi and F. Mojtabai, Phys. Rev. A {\bf 26},  1168  (1982).

\bibitem{carm82}
B. Carmeli and A. Nitzan, Phys. Rev. Lett. {\bf 49},  423  (1982).

\bibitem{hern99b}
R. Hernandez and F.~L. Somer, J. Phys. Chem. B {\bf 103},  1070  (1999).

\bibitem{hern99d}
F.~L. Somer and R. Hernandez, J. Phys. Chem. B {\bf 104},  3456  (2000).

\bibitem{haynes94}
G.~R. Haynes, G.~A. Voth, and E. Pollak, J. Chem. Phys. {\bf 101},  7811
  (1994).

\bibitem{art2}
E. Hershkovitz, J. Chem. Phys. {\bf 108},  9253  (1998).

\bibitem{chan78}
D. Chandler, J. Chem. Phys. {\bf 68},  2959  (1978).

\bibitem{frish93a}
A.~M. Frishman and E. Pollak, J. Chem. Phys. {\bf 98},  9532  (1993).

\bibitem{pollak86b}
E. Pollak, J. Chem. Phys. {\bf 85},  865  (1986).

\bibitem{hern99e}
R. Hernandez, J. Chem. Phys. {\bf 110},  7701  (1999).

\bibitem{kloe92}
P.~E. Kloeden and E. Platen, {\em Numerical Solution of Stochastic Differential
  Equations} (Springer--Verlag, New York, 1992).

\bibitem{numrec}
W.~H. Press, B.~P. Flannery, S.~A. Teukolsy, and W.~T. Vetterling, {\em
  Numerical Recipes} (Cambridge University Press, Cambridge, UK, 1988).

\bibitem{allen87}
M.~P. Allen and D.~J. Tildesley, {\em Computer Simulations of Liquids} (Oxford,
  New York, 1987).

\bibitem{chandler}
D. Chandler, {\em Introduction to Modern Statistical Mechanics} (Oxford
  University Press, New York, 1987).

\end{thebibliography}

\end{document}